# Critical Percolation Probabilities for the Next-Nearest-Neighboring Site Problems on Sierpinski Carpets

H. B. Nie, B. M. Yu

*Department of Physics, Huazhong University of Science and Technology, Wuhan 430074, China*

K. L. Yao

*Center of Theoretical Physics, CCAST (World Laboratory), Beijing 100080, and Department of Physics, Huazhong University of Science and Technology, Wuhan 430074, China*

## Abstract

In this paper, we compute the next-nearest-neighboring site percolation (Connections exist not only between nearest-neighboring sites, but also between next-nearest-neighboring sites.) probabilities $P_c$ on the two-dimensional Sierpinski carpets, using the translational-dilation method and Monte Carlo technique. We obtain a relation among $P_c$, fractal dimensionality $D$ and connectivity $Q$. For the family of carpets with central cutouts, $(1-P_c)/(1-P_c^s) = (D-1)^{1.60}$, where $P_c^s = 0.41$, the critical percolation probability for the next-nearest-neighboring site problem on square lattice. As $D$ reaches 2, $P_c = P_c^s = 0.41$, which is in agreement with the critical percolation probability on 2-d square lattices with next-nearest-neighboring interactions.

## I. Introduction

In recent years, the study of fractal geometries has been receiving great attention [1-12]. Gefen *et al.* [3-5] discussed the self-similar lattices with a finite order of ramification $R$ and with $R = \infty$. In particular, they applied the connectivity $Q$, fractal dimensionality $D$ to characterize the Sierpinski carpet families with $R = \infty$, and found that discrete-symmetry spin models on those lattices undergo a phase transition at $T_c > 0$.

Gefen *et al.* [5] constructed an approximate renormalization-group scheme for the Ising model in the Sierpinski carpet family and found the dependence of $T_c$ and the critical exponents on the geometrical factors. Riera [7] analysed the critical behaviour of the ferromagnetic *q*-state Potts models on Sierpinski carpets, and used an approximate Migdal-Kadanoff real space renormalization-group scheme to find the dependence of the critical temperature and the critical exponents on the number of states $q$ as well as on geometrical parameters.

Yu and Yao [11] computed the nearest-neighboring percolation probabilities $P_c$ on the two-dimensional Sierpinski carpets, and found a relation among $P_c$, fractal dimensionality $D$ and connectivity $Q$.

Qu, Yao and Yu [12] used the renormalization-group approach and Monte-Carlo simulation to treat the problem of percolation on two-dimensional square lattices with next-nearest-neighboring interactions. The critical probability $P_c^{"}$ and critical exponents α, β, γ were obtained.

In this paper we use the translational-dilation method (TDM) proposed Yu and Yao [11], and Monte Carlo technique to compute $P_c$ for the next-nearest-neighboring site percolation problems (Connections exist not only between nearest-neighboring sites, but also between next-nearest-neighboring sites.) probabilities on Sierpinski carpet families. We find that $P_c$ depends on the fractal dimensionality $D$ and connectivity $Q$. Especially, we obtain a relation among $P_c$ and $D$, and $Q$. We also find that $P_c$ in the family of carpets with central cutouts is greater than that of the family with evenly scattered cutouts. Compared with the nearest-neighboring percolation probabilities $P_c^{'}$ on the two-dimensional Sierpinski carpets [11], for the same $D$ and $Q$ (same $b$ and $l$), we have $P_c < P_c^{'}$. Compared with the next-nearest-neighboring site probabilities $P_c^{"}$ on the two-dimensional square lattice [12], we have $P_c > P_c^{"}$.

## II. Method of Construction for the Sierpinski Carpet Lattices

For the convenience of computing $P_c$ on the Sierpinski carpet family numerically, one needs to transfer the geometry problem for Sierpinski carpet lattices into a digital problem. In this paper we use the translational-dilation method (TMD) to construct a Sierpinski carpet lattice. Figure 1a (central cutout) and 1b (scattered cutout) show examples of self-similar fractal lattice called Sierpinski carpets, which are formed by the TDM. The recursion formulas for TDM are as follows: We first assign an initial value for a square lattice, for example

$$M(I, J) = 1 \qquad (1)$$

Where $1 \leq I \leq 2$, $1 \leq J \leq 2$, $M(I, J)$ is an element of two-dimensional array, $I$ and $J$ represent the row and column on a carpet lattice respectively, see Fig. 3a.

Then, the sites are translated towards the below according to

$$M(I,J) = M(I - b^{n-1}, J), \qquad (2)$$

where

$$T \times b^{n-1} + 2 \leq I \leq (T+1)b^{n-1} + 1,$$

$1 \leq T \leq b-1$, $1 \leq J \leq b^{n-1} +1$, $1 \leq n \leq N$, $n$ is the serial number of translation, $N$ is the stages of Sierpinski carpets (see Fig. 3b, Fig. 2a-c). $T$ is a translational parameter. Equation (2) is also adequate for the scattered cutout carpet.

And then the known sites are translated toward the right by

$$M(I,J) = M(I, J - b^{n-1}),  \qquad (3)$$

where $1 \leq I \leq (b-l)b^{n-1}/2 +1$ and $(b+l)b^{n-1} +1 \leq I \leq b^n +1$, $b^{n-1} +2 \leq J \leq b^n +1$; for the scattered cutout carpet, where $(T-1)b^{n-1} +1 \leq I \leq T \times b^{n-1} +1$, $T$ is an odd number less than or equal to $b$, $b^{n-1} +2 \leq J \leq b^n +1$, see Fig. 3c.

Finally, the sites are translated towards the below by

$$M(I,J) = M(I - b^{n-1}, J),  \qquad (4)$$

where $(b-l)b^{n-1}/2 +2 \leq I \leq (b+l)b^{n-1}/2$ (if $l = 1$ and $n = 1$ this inequality is eliminated), $b^{n-1} +2 \leq J \leq (b-l)b^{n-1}/2 +1$ (if $b- l = 2$, this inequality is eliminated) as well as $(b+l)b^{n-1}/2 +1 \leq J \leq b^n +1$; for the scattered cutout carpet, where $(T-1)b^{n-1} +2 \leq I \leq T \times b^{n-1}$ ($T$ is an even number less than $b$); $T' \times b^{n-1} +1 \leq J \leq (T' +1)b^{n-1} +1$ ($T'$ is an even number less than $b$). If $n = 1$, the calculation of Eq. (4) is eliminated.

If $n$ is increased by 1, the dilation of the lattice is generated, then the translational calculation processes are repeated, and so on. Finally the requested lattice for the Sierpinski carpets is formed (see Fig. 2).

For the cutouts of a carpet lattice, the computer can automatically assign the zero to the cutout sites (see the circles in Fig. 3d).

### III.    Site Occupation and Cluster Calculation on the Sierpinski Carpet

Sites on the Sierpinski Carpet are randomly occupied to rows ($I$) and columns ($J$). The probabilities $P_i$ ($i = 1, 2, 3, …$) are separately compared with the pseudo-random number $y(0 \sim 1)$. If $P_i < y$, the site ($I, J$) is not occupied and assigned to be zero otherwise, a positive integer is reassigned to the site ($I, J$). If the site ($I, J$) is occupied, nearest-neighboring sites (see sites ($I, J - 1$) and ($I - 1, J$) in Fig. 4) and next-nearest-neighboring sites (see sites ($I - 1, J - 1$) and ($I - 1, J + 1$) in Fig. 4) already compared with $y$ are necessarily investigated to see whether they are occupied or not, if so it belongs to the cluster composed of neighboring sites the serial number of the cluster is assigned to the site ($I, J$), the size of the cluster is then increased by unity; if not, the site (*I*, *J*) forms a new cluster (its size is 1), and a new serial number which is greater than the last one by 1 is assigned to the new cluster, and so on. Two clusters coalesce into a greater cluster if two clusters are joined by (at least) a bond (including a bond which joins two next-nearest-neighboring sites) which joining two sites that belong to different clusters. If two clusters coalesce into a new greater cluster, the serial number (e.g. no. 7) of sites of the latter cluster is then changed to be the serial number (e.g. no. 2) of sites of the former cluster. The size of the new cluster is then equal to the sum of sites of two clusters, and finally zero is assigned to the latter cluster size.

### IV. Critical Percolation Probabilities $P_c$ and Discussion

We again applied the modified second moment

$$M_i = \frac{\sum_i S_i^2}{\left(\sum S_i\right)^2} \tag{5}$$

defined by Dean [9], where $S_i$ ($i = 1, 2, 3, \ldots$) are the sizes of clusters and the summation runs over all the lattice clusters.

Then, $P_c$ is taken to be that value of $P$ at which $\Delta M / \Delta P$ is maximum, $\Delta$ indicating an increament of one step. In the present work we choose $\Delta P$ to be a constant (e.g. $\Delta P = 0.01$) and set $P_i$ ($i = 1, 2, 3, \ldots$), is calculated from Eq. (5). The critical percolation probability for site problems on Sierpinski carpets is

$$P_c = \frac{P_i + P_{i+1}}{2} \tag{6}$$

when $(M_{i+1} - M_i)/\Delta P$ reaches the maximum.

From the computation of the cluster-size distribution the critical percolation probabilities can be obtained according to Eq. (6). The results for various *b*, *l* of Sierpinski carpets are summarized in Tables 1 and 2, where the averages are taken over 10 runs.

It can be seen, from Tables 1 and 2, that for the same *b* the critical percolation probabilities $P_c$ increases with *l*, and $P_c$ increases with *D* decreasing and *Q* decreasing.

From our numerical calculations we obtain the approximate relations, which are shown in Figs 5, 6 and 7.

For the family of carpets with central cutouts:

$$P_c = 1 - (1 - P_c^s)(D - 1)^{1.60} \tag{7}$$

or

$$\frac{1-P_c}{1-P_c^s} = (D-1)^{1.60} \tag{8}$$

and

$$P_c = 0.32Q^2 - 0.92Q + 1.01 \tag{9}$$

with the remainder error (defined by $e_r = \{[1/(k-2)]\sum_{i=1}^{k}\sigma_{ei}^2\}^{1/2}$, here $k = 22$) 0.008 and 0.015 respectively (see Table 1). Here $P_c^s$ is the critical percolation probability on the next-nearest-neighboring square lattice, and $P_c^s = 0.41$.

For the family of carpets with evenly cutouts:

$$P_c = 0.36Q^2 - 0.88Q + 0.92 \tag{10}$$

with the remainder error 0.008 (see Table 2).

From Eq. (7) and Fig. 5, we can find, as $D$ reaches 2.0, for the family of carpets with central cutouts, $P_c = P_c^s = 0.41$, which is in agreement with the critical percolation probability on 2-$d$ square lattices with next-nearest-neighboring interactions [1].

From Eq. (9) and Fig. 7, we can find, as $Q$ reaches 1.0 for the family of carpets with central cutouts, $P_c = P_c^s = 0.41$, which is in agreement with the critical percolation probability on 2-$d$ square lattices with next-nearest-neighboring interactions [1].

From Eq. (10) and Fig. 7, we can find, for the family of carpets with evenly scattered cutouts, as $Q$ reaches 1.0, $P_c = 0.40 \pm 0.01$, which is also in agreement with the critical percolation probability on 2-$d$ square lattices with next-nearest-neighboring interactions [1].

From Tables 1-2 and Figs 5-8, one easily finds that for the same $D$ or $Q$ (same $b$ and $l$), $P_c$ in the family of carpets with central cutouts is greater than those of the family with evenly scattered cutouts. This is expected, since they belong to two different universality classes [11].

Compared with the nearest-neighboring percolation probabilities $P_c^{'}$ on the two-dimensional Sierpinski carpets [11], for the same $D$ or $Q$ (same $b$ and $l$), we have $P_c < P_c^{'}$. This is reasonable, since for the next-nearest-neighboring site percolation, connections exist not only between nearest-neighboring sites, but also between next-nearest-neighboring sites, thus $P_c$ becomes smaller.

Compared with the site percolation probabilities $P_c^{"}$ on the two-dimensional square lattice [12], we have $P_c > P_c^{"}$. This is also reasonable, since for the Sierpinski carpet site percolation, the connectivity is smaller than that of the square lattice site percolation, thus $P_c$ is larger than $P_c''$.

For the family of carpets with central cutouts, and for the nearest-neighboring site percolation, Yu and Yao [11] obtained,

$$\frac{1-P_c}{1-P_c^s} = (D-1)^{2.35}, \qquad (11)$$

where $P_c^s = 0.593$ is the critical percolation probability on the nearest-neighboring square lattices. For the next-nearest-neighboring site percolation on the Sierpinski carpets, we obtain

$$\frac{1-P_c}{1-P_c^s} = (D-1)^{1.60}, \qquad (12)$$

where $P_c^s = 0.41$ is the critical percolation probability on the next-nearest-neighboring square lattices.

It is interesting that the exponent 1.60 for the next-nearest-neighboring site percolation on the Sierpinski carpets is different from the exponent 2.35 for the nearest-neighboring site percolation on the Sierpinski carpets. Until now, no knowledge about the physical origin of exponents 1.60 and 2.35 is reported. It is an open question and needs further study.

### V. Conclusion

In this paper, we compute the next-nearest-neighboring site percolation probabilities $P_c$ on the two-dimensional Sierpinski carpets, using, using the translational-dilation method and Monte Carlo technique. We obtain a relation among $P_c$, fractal dimensionality $D$ and connectivity $Q$. For the family of carpets with central cutouts, $(1-P_c)/(1-P_c^s) = (D-1)^{1.60}$, where $P_c^s$ is the critical percolation probability for the next-nearest-neighboring square lattices. As $D$ reaches 2, $P_c = P_c^s = 0.41$, which is in agreement with the critical percolation probability on 2-d square lattice with next-nearest-neighboring interactions [1].

Interestingly, we noticed that the exponent 1.60 for the next-nearest-neighboring site percolation on the Sierpinski carpets is different from the exponent 2.35 for the nearest-neighboring site percolation on the Sierpinski carpets. Until now, no knowledge about the physical origin of exponents 1.60 and 2.35 is reported. It is an open question and needs further study.

**Table 1.** Critical percolation probabilities $P_c$ for the next-nearest-neighboring site problems on Sierpinski carpets with central cutouts $(l \times l)$. Here $\sigma_e$ is an error between experiment value and predicted value from Eq. (7) or Eq. (9), $\sigma$ is the standard error.

| b | l | N | D | Q | Experiment Value $P_c \pm \sigma$ | Predicted value from Eq. (7) $P_c$ $\sigma_e$ | Predicted value from Eq. (9) $P_c$ $\sigma_e$ |
|---|---|---|---|---|---|---|---|
| 13 | 11 | 2 | 1.509 | 0.270 | 0.816 ± 0.005 | 0.800 – 0.016 | 0.785 – 0.031 |
| 11 | 9 | 2 | 1.588 | 0.289 | 0.788 ± 0.007 | 0.781 – 0.007 | 0.771 – 0.017 |
| 9 | 7 | 2 | 1.577 | 0.316 | 0.748 ± 0.003 | 0.755 + 0.007 | 0.751 + 0.003 |
| 7 | 5 | 2 | 1.633 | 0.356 | 0.703 ± 0.006 | 0.716 + 0.013 | 0.723 +0.020 |
| 5 | 3 | 3 | 1.723 | 0.431 | 0.649 ± 0.005 | 0.648 – 0.001 | 0.673 + 0.024 |
| 13 | 9 | 2 | 1.746 | 0.541 | 0.631 ± 0.004 | 0.631 + 0.000 | 0.606 – 0.025 |
| 11 | 7 | 2 | 1.784 | 0.578 | 0.601 ± 0.004 | 0.600 – 0.001 | 0.585 – 0.016 |
| 9 | 5 | 2 | 1.832 | 0.631 | 0.577 ± 0.004 | 0.560 – 0.017 | 0.557 – 0.020 |
| 15 | 9 | 2 | 1.835 | 0.662 | 0.541 ± 0.005 | 0.558 + 0.017 | 0.541 + 0.000 |
| 13 | 7 | 2 | 1.867 | 0.699 | 0.524 ± 0.003 | 0.530 + 0.006 | 0.523 – 0.001 |
| 7 | 3 | 2 | 1.896 | 0.712 | 0.508 ± 0.008 | 0.505 – 0.003 | 0.517 + 0.009 |
| 11 | 5 | 2 | 1.903 | 0.747 | 0.497 ± 0.003 | 0.499 + 0.002 | 0.501 + 0.004 |
| 15 | 7 | 2 | 1.909 | 0.768 | 0.491 ± 0.003 | 0.494 + 0.003 | 0.492 + 0.001 |
| 13 | 5 | 2 | 1.938 | 0.811 | 0.464 ± 0.004 | 0.467 + 0.003 | 0.474 + 0.010 |
| 9 | 3 | 2 | 1.946 | 0.816 | 0.463 ± 0.004 | 0.460 – 0.003 | 0.472 + 0.009 |
| 15 | 5 | 2 | 1.957 | 0.850 | 0.446 ± 0.002 | 0.450 + 0.004 | 0.459 + 0.013 |
| 11 | 3 | 2 | 1.968 | 0.867 | 0.442 ± 0.003 | 0.440 – 0.002 | 0.453 + 0.011 |
| 5 | 1 | 3 | 1.975 | 0.861 | 0.438 ± 0.003 | 0.433 – 0.005 | 0.455 + 0.017 |
| 13 | 3 | 2 | 1.979 | 0.898 | 0.436 ± 0.003 | 0.430 – 0.006 | 0.442 + 0.006 |
| 7 | 1 | 2 | 1.989 | 0.921 | 0.430 ± 0.005 | 0.420 – 0.010 | 0.434 + 0.004 |
| 9 | 1 | 2 | 1.994 | 0.946 | 0.417 ± 0.003 | 0.416 – 0.001 | 0.426 + 0.009 |
| 13 | 1 | 2 | 1.998 | 0.969 | 0.415 ± 0.003 | 0.412 – 0.003 | 0.419 + 0.004 |

**Table 2.** Critical percolation probabilities $P_c$ for the next-nearest-neighboring site problems on Sierpinski carpets with evenly scattered cutouts $(l \times l)$. Here $\sigma_e$ is an error between experiment value and predicted value from Eq. (10), $\sigma$ is the standard error.

| b | l | N | D | Q | Experiment Value $P_c \pm \sigma$ | Predicted value from Eq. (10) $P_c$ $\sigma_e$ |
|---|---|---|---|---|---|---|
| 5 | 2 | 3 | 1.892 | 0.683 | 0.490 ± 0.002 | 0.487 – 0.003 |
| 7 | 3 | 2 | 1.896 | 0.712 | 0.463 ± 0.003 | 0.476 – 0.013 |
| 9 | 4 | 2 | 1.900 | 0.734 | 0.461 ± 0.002 | 0.468 + 0.007 |
| 11 | 5 | 2 | 1.904 | 0.747 | 0.459 ± 0.003 | 0.464 + 0.005 |
| 13 | 6 | 2 | 1.907 | 0.759 | 0.456 ± 0.003 | 0.459 + 0.003 |
| 15 | 7 | 2 | 1.909 | 0.768 | 0.455 ± 0.002 | 0.456 + 0.001 |
| 17 | 8 | 2 | 1.912 | 0.776 | 0.447 ± 0.002 | 0.454 + 0.007 |

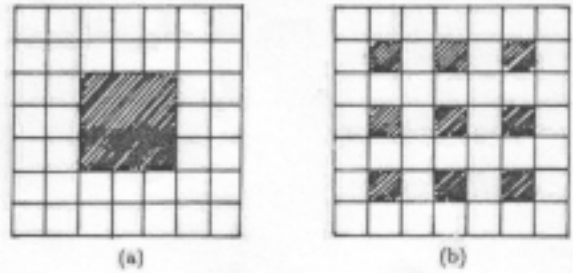

Fig. 1. Sierpinski carpets with $R = \infty, b = 7, l = 3$ after one step ($n = 1$) in the construction of the lattices according to the TDM. The lattice sites are denoted by black dot. The fractal dimensionality $D = \ln(b^2 - l^2)/\ln b$, the connectivity $Q = \ln(b - l)/\ln b$, figures 1a and 1b have the same $D$ and $Q$, but 1a for large lacunarity with central cutouts, and 1b for small lacunarity with evenly scattered cutouts.

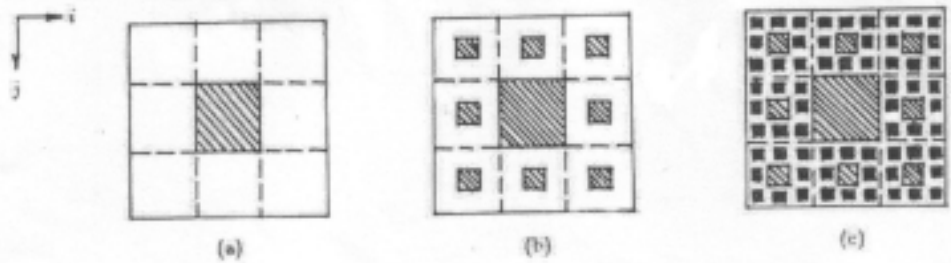

Fig. 2. Three stages of a Sierpinski carpet and division of regions. (a). $N = 1$-stage Sierpinski carpet, the side length $b = 12$. (b). $N = 2$-stage Sierpinski carpet (the side length is shortened 3 times). (c). $N = 3$-stage Sierpinski carpet (the side length is shortened 9 times).

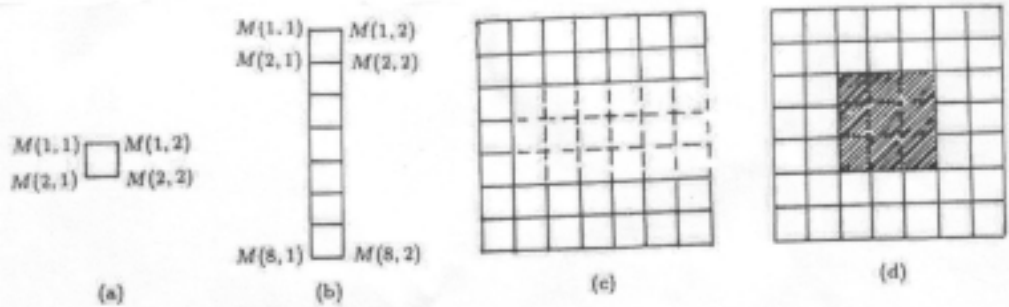

Fig. 3. The construction of a carpet lattice with central cutouts ($b = 7, l = 3, n = 1$).

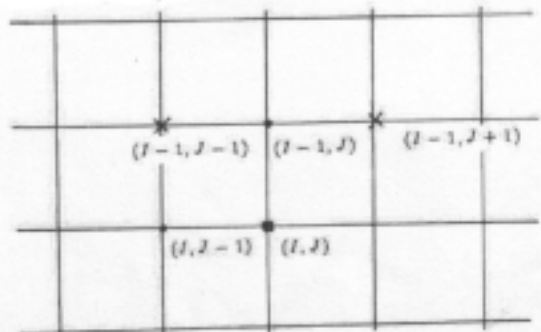

Fig. 4. Nearest-neighbouring sites and next-nearest-neighbouring sites.

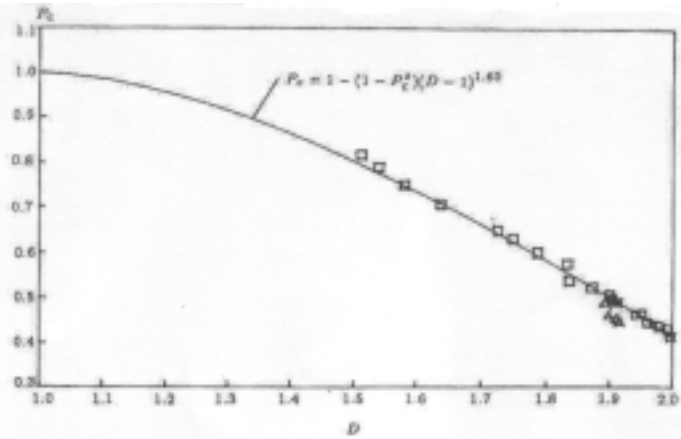

Fig. 5. The relations of $P_c$ and $D$. $\square$ represents those for carpets with central cutouts. $\triangle$ represents those for carpets with evenly scattered cutouts.

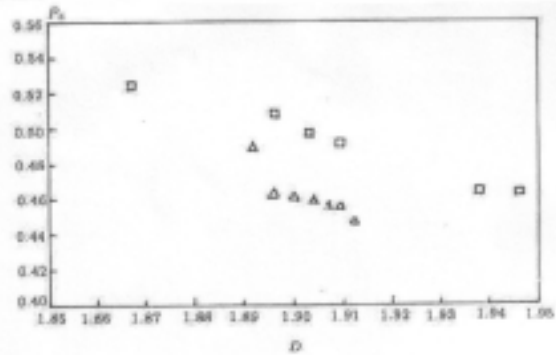

Fig. 6. A part of Fig. 5.

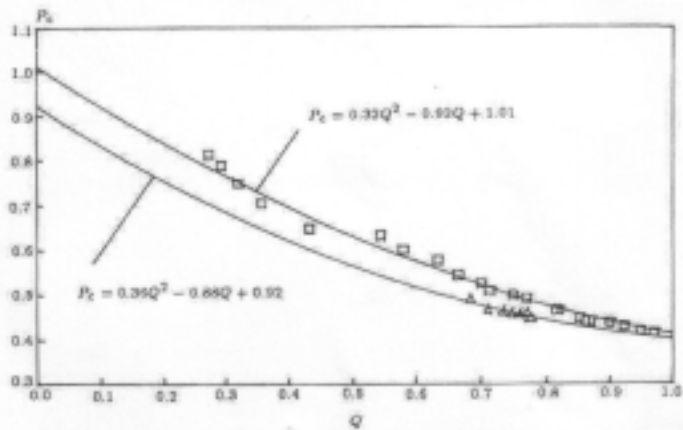

Fig. 7. The relations of $P_c$ and $Q$. $\square$ represents those for carpets with central cutouts. $\triangle$ represents those for carpets with evenly scattered cutouts.

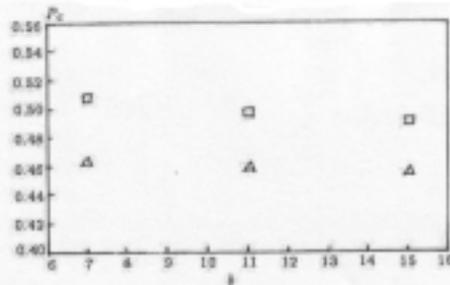

Fig. 8. For the same $b$ and $l$, $P_c$ comparison, $\square$ – central cutout carpets, $\triangle$ – evenly scatterrd cutout carpets.